\documentclass[a4paper,12pt]{article}
\pdfoutput=1 % if your are submitting a pdflatex (i.e. if you have
\usepackage{jcappub} % for details on the use of the package, please
\usepackage[T1]{fontenc}
\usepackage{graphicx}
\usepackage{epsf}
\usepackage{bm,color}
\usepackage{float}
\usepackage{amsmath}
\usepackage{amsfonts}
\usepackage{amssymb}
\usepackage{epstopdf}
\usepackage{natbib}%
\usepackage{subcaption}
\usepackage{hyperref}
\hypersetup{
	bookmarks=true,         % show bookmarks bar?
	unicode=false,          % non-Latin characters in Acrobat’s bookmarks
	pdftoolbar=true,        % show Acrobat’s toolbar?
	pdfmenubar=true,        % show Acrobat’s menu?
	pdffitwindow=false,     % window fit to page when opened
	pdfstartview={FitH},    % fits the width of the page to the window
	pdftitle={MHI},    % title
	pdfauthor={Author},     % author
	pdfsubject={Subject},   % subject of the document
	pdfcreator={Creator},   % creator of the document
	pdfproducer={Producer}, % producer of the document
	pdfkeywords={keyword1} {key2} {key3}, % list of keywords
	pdfnewwindow=true,      % links in new window
	colorlinks=true,       % false: boxed links; true: colored links
	linkcolor=red,          % color of internal links
	citecolor=cyan,        % color of links to bibliography
	filecolor=magenta,      % color of file links
	urlcolor=blue,           % color of external links
	linktocpage=true
}
\providecommand{\U}[1]{\protect\rule{.1in}{.1in}}
\newcommand{\be}{\begin{equation}}
	\newcommand{\ee}{\end{equation}}

\newcommand{\mincir}{\raise
	-3.truept\hbox{\rlap{\hbox{$\sim$}}\raise4.truept\hbox{$<$}\ }}
\newcommand{\magcir}{\raise
	-3.truept\hbox{\rlap{\hbox{$\sim$}}\raise4.truept\hbox{$>$}\ }}

\def\bea{\begin{eqnarray}}
	\def\eea{\end{eqnarray}}
\def\ba{\begin{array}}
	\def\ea{\end{array}}

\def\beq{\begin{equation}}
	\def\eeq{\end{equation}}

\newcommand{\eq}{Eq.\eqref}
\newcommand{\fig}{Fig.\ref}
\usepackage{ulem}

\newcommand{\phicmb}{\phi_{\rm CMB}}
\newcommand{\phiend}{\phi_{\rm end}}

\newcommand{\alphaa}{2^{\frac{3}{4}}\alpha^{\frac{3}{2}}}

	\title{The fate of Quasi-Exponential inflation in the light of ACT-DR6}
	\author{Barun Kumar Pal}
	\emailAdd{terminatorbarun@gmail.com}
	\affiliation{Netaji Nagar College For Women, Kolkata-700092, West Bengal, India}
	\abstract {We have revisited quasi-exponential model of inflation in the light of recent ACT-DR6 and Planck data along with latest constraint on the amplitude of primordial gravitational waves. For our analysis we have followed Mukhanov approach for inflationary equation-of-state employing Hamilton-Jacobi formulation. We find that the model is capable of mimicking  latest Planck results by providing excellent fit to scalar spectral index and its running. Not only that, amount of primordial gravitational waves is also within the present observational bound, $r<0.032$. In addition to that, when the combination of  ACT-DR6 and Planck joint with BICEP/Keck 2018 data is taken into account inflationary predictions from quasi-exponential model are  in excellent agreement. The model also yields sublime fit to the result of joint analysis of ACT-DR6, Planck joint with BICEP/Keck 2018 and DESI-Y1 data.  The futuristic CMB missions LiteBIRD, Simons Observatory and CMB-S4 are promising to detect primordial gravitational waves if $r\geq0.003$. We have also forecasted inflationary predictions from quasi-exponential model of inflation assuming  the sensitivities of LiteBIRD, Simons Observatory and CMB-S4 along with  combination of LiteBIRD and  CMB-S4. We found that in each case quasi-exponential inflation may render excellent fit provided $r\geq10^{-2}$. However,  non-detection  of primordial gravitational waves by LiteBIRD, Simons Observatory and CMB-S4 will potentially rule out quasi-exponential model.   
}
	
\begin{document}
	\maketitle
	\flushbottom	
	
	%%%%%%%%%%%%%%%%%%%%%%%%%%%%%%%%%%%%%%%%%%%%%%%%%%%%%%%%%%%%%%%%%%%%%%%%%%%%%%%%%%%%

%%%%%%%%%%%%%%%%%%%%%%%%%%%%%%%%%%%%%%%%%%%%%%%%%%%%%%%%%%%%%%%%%%%%%%%%%%%%%%%%%%%%
\section{Introduction} \label{sec1}
Cosmic inflation \cite{starobinsky1978, guth1982, starobinsky1982} has remained the most omnipotent instrument to retaliate critique   and inadequacy of the standard big bang theory. Since its inception many different models of inflation has been proposed by the cosmologists owing to the wide observational window allowed by contemporary cosmological probes.  With the advent of precise data from different probes in the likes of WMAP, Planck, SDSS \cite{akrami2020planck, aghanim2020planck, ade2014planck, spergel2007}  has already ruled out numerous inflationary models. Very recent data from Atacama Cosmology Telescope (ACT) \cite{louis2025atacama,  calabrese2025atacama} has anticipated higher value of the scalar spectral index which may have the potential to subside the number of viable inflationary models by narrowing the observational window. The upcoming CMB missions in the likes of BICEP2/Keck \cite{ade2018constraints}, CMB-S4 \cite{abazajian2016cmb}, LiteBIRD \cite{matsumura2014mission}, Simons Observatory \cite{SimonsObservatory:2018} are expected to reduce further the number of observationally viable  inflationary models by surveying the primordial gravitational waves up to $r\sim\mathcal{O}(10^{-3})$.

During its prolonged existence the literature of cosmic inflation has struck it rich. But still it has remained as a paradigm since  a specific compelling model of inflation is yet to be separated out from the spectrum of observationally viable inflationary models. The upper-bound on tensor-to-scalar ratio has been pushed back to $r<0.032$ \cite{tristram2022improved}. The analysis of Planck-2018 data has put  stringent constraint on the scale dependence of scalar curvature perturbation, $n_{_S}=0.9649\pm 0.0042$ \cite{aghanim2020planck,akrami2020planck} and very recent analysis from Planck data has brought forth $n_{_S}=0.9690\pm 0.0035$ \cite{tristram2024}. However the latest ACT-DR6 data has given a stronger indication for higher value of the spectral index, $n_{_S}=0.9666\pm0.0077$. Joint analysis of Planck and ACT-DR6 data has come up with $n_{_S}=0.9709\pm0.0038$ \cite{louis2025atacama, calabrese2025atacama}. Further when Planck and ACT-DR6 data  are combined with DESI-Y1   yields  $n_{_S}=0.9743\pm0.0034$ \cite{adame2025desi1, adame2025desi}, which is touch closer to unity. Thus inflationary models anticipating lower amplitude of primordial gravitational waves and higher value for  scalar spectral index will now have the edge over the others from the observational point of view. This is exactly what quasi-exponential  model \cite{barunquasi} of inflation does. The Hubble parameter here being near exponential in nature  seeds almost scale invariant  scalar curvature perturbation. Though being large field model quasi-exponential inflation is expected to generate higher amount of primordial gravitational  waves, but that can be accounted for by keeping the window open for spectral index towards scale invariance  as we shall see later on.

In this article we have used Hamilton-Jacobi formulation \cite{muslimov1990, salopek1990}, where the Hubble parameter, $H$, is treated as the fundamental quantity in contrast to the traditional approach of inflationary cosmology where we need  to specify a particular form of the scalar field potential, $V$, to analyze inflationary dynamics. Main objective of this approach is to incorporate different types of inflationary models irrespective of slow-roll approximations \cite{linde1982, albrecht1982, liddle1994}.  As slow-roll approximation is not the only way to go for inflationary dynamics and solutions beyond slow-roll approximations has also been found \cite{wands1996}. Being first order in nature, these equations are easily tractable to explore the underlying physics. One of the main benefit of this approach being it is more accurate than the usual slow-roll method as it also takes into account the effect from  the kinetic term present in inflationary dynamics. Considering the precision level achieved by the present day detectors, the inflationary predictions
should now be very precise to go with the latest observations.

The production of nearly scale invariant scalar curvature perturbation in  quasi-exponential inflation has almost ruled it out from the modern inflationary literature. With the recent findings by the latest ACT data in combination with Planck-2018 results, QEI has become alive once again. In this work we shall see that, it  not only addresses the higher spectral index and smaller tensor-to-scalar ratio, but it can also explain futuristic CMB missions in the likes of LiteBIRD, Simons Observatory and CMB-S4 as well,   provided $r\sim\mathcal{O}(10^{-2})$.

In this work we have confronted quasi-exponential inflation (QEI henceforth) with latest data from ACT. Previous investigation of QEI relied on earlier CMB data set mainly WMAP and Planck. The current study focuses on  newly released  ACT-DR6 data set as well as latest Planck results. By combining ACT-DR6 with Planck-2018, we test QEI predictions with latest observational constraints thereby providing updated assessment of the model’s viability. This is also the first attempt to constraint QEI in the light of ACT-DR6. We have utilized constraint from the joint analysis of Planck and ACT-DR6 data along with the combination of Planck, ACT-DR6 and DESI-Y1 data employing Hamilton-Jacobi formulation following Mukhanov parametrization \cite{mukhanov2013quantum} of inflationary equation-of-state. We have further utilized predictions for tensor-to-scalar ratio  from forthcoming CMB missions in the likes of LiteBIRD, Simons Observatory and CMB-S4 to constrain quasi-exponential inflation. In the process we are also able to put strong constraint on the model parameter.

\section{Quasi Exponential Inflation}\label{qhj}
The inflationary Hubble parameter that we are interested in here has the following form \cite{barunquasi}
\beq
H(\phi)={H_0} \exp\left[{\frac{\alpha\phi{\rm M^{-1}_{P}}}{ (\phi {\rm M^{-1}_{P}}+1)}}\right],
\eeq
where $\alpha$ is a dimensionless free parameter. This model was first introduced  in \cite{barunquasi} and later investigated in \cite{videla2017}. Here we shall thoroughly examine the pros and cons of this model in the view of latest ACT-DR6 data \cite{calabrese2025atacama, louis2025atacama} in combination with DESI-Y1\cite{adame2025desi1, adame2025desi} and latest Planck result \cite{ akrami2020planck, aghanim2020planck, ade2021improved, tristram2022improved, tristram2024} within the framework of Mukhanov parametrization of inflationary equation of state \cite{mukhanov2013quantum} employing Hamilton-Jacobi formulation \cite{salopek1990, muslimov1990,liddle1994, kinney1997,lidsey1997, barunmhi, barunquasi,barun2018mutated, baruneos2023, barunmeos2025}. Within this formulation the exact end of inflation, for the model under consideration, is given by \cite{barunquasi}
 \beq
\phiend {\rm M^{-1}_{P}}=-1+2^{\frac{1}{4}}\sqrt{\alpha}.
\eeq
Restricting ourselves to the positive values of the inflaton we get $\alpha \geq\frac{1}{\sqrt{2}}$. The upper bound on this model parameter may be fixed from the observational bound, which we shall comeback later.

The expression for the number of e-foldings then turns out to be 
\bea
\text{N}&=&\frac{1}{6 \alpha }\left(1+\phi M_P^{-1}\right)^3 - \frac{\alphaa}{6 \alpha}.
\eea
The above equation can be inverted easily to have scalar field as a function of e-folding, 
\beq
\phi M_P^{-1}=\sqrt[3]{6\alpha \text{N}+\alphaa}-1.
\eeq
%The so-called slow-roll parameter $\epsilon_{_H}$ may also be re-written as a function of e-foldings, 
%\beq
%\epsilon_{_H}=\frac{2\alpha^2}{(6\alpha \text{N}+\Phi^3_{\rm end})^{4/3}}.
%\eeq
%%%%%%%%%%%%%%%%%%%%%%%%%%%%%%%%%%%%%%%%%%%%%%%%%%%%%%%%%%%%%%%%%%%%%%%%%%%%%%%%%%%%%%%%%%%%%%%%%%%%
%%%%%%%%%%%%%%%%%%%%%%
\section{Mukhanov Parametrization}
In Mukhanov parametrization we express inflationary equation-of-state as a function of number of e-foldings, $N$. The main goal was to develop a model independent framework for investigation and confrontation of cosmic inflation with recent observations \cite{mukhanov2013quantum}. But this need not be true as we shall see that specifying a particular equation-of-state  directly leads to choosing a specific inflationary potential \cite{martin2016observational}. However the equation-of-state formalism provides a simple yet  elegant way to confront inflationary observables with the observations as we shall see. 

The condition for inflation may  be put forward through the equation-of-state parameter,  $\omega(\phi)\equiv\frac{p}{\rho}<-1/3$, which within quasi-exponential inflation turns out to be
\bea\label{ephn}
1+\omega(N)\equiv\frac{4}{3}\frac{\alpha^2}{(6\alpha \text{N}+\alphaa)^{4/3}}<\frac{2}{3}.
\eea
Though we are considering a particular model of inflation,   the above form of the  equation-of-state closely mimics Mukhanov Parametrization \cite{mukhanov2013quantum}. 
With the help of  \eq{ephn} we can easily rewrite the Hubble parameter as a function of the e-foldings, 
\beq\label{hn}
H(N)={H_0}e^{\alpha\left(1-\frac{1}{\sqrt[3]{6\alpha \text{N}+\alphaa}}\right)}.
\eeq
We can also derive the exact expression of the associated inflationary potential using Hamilton-Jacobi equation in the context of inflation, which turns out to be
\beq
V(\phi)={H_0}^2{\rm M_{P}}^2\left[3-\frac{2\alpha^2}{(1+\phi{\rm M^{-1}_{P}})^4}\right] e^{2\alpha\left(1-\frac{1}{1+\phi{\rm M^{-1}_{P}}}\right)}.
\eeq
This potential remains positive for feasible values of the model parameter. For large field values $\phi\gg M_P$, the correction term $(1+\phi{\rm M^{-1}_{P}})^{-4}$ becomes negligible and the potential approaches a constant plateau, $V(\phi)\sim 3{H_0}^2{\rm M_{P}}^2e^{2\alpha}$. This flat region allows scalar field to evolve slowly and helps in generating almost scale invariant primordial perturbations. As inflation goes on the field value decreases, the correction term becomes significant reducing the potential and producing minima at a finite field value $\phi_{\rm{min}}$, which depends on the model parameter $\alpha$. As we have restricted ourselves to the positive value for the scalar field, $\phi_{\rm{min}}=0$  and hence $V(\phi_{\rm{min}})>0$. 
After the end of inflation, the scalar field rolls down from the plateau and settles near this minimum. The field then undergoes damped oscillations about the minimum. These oscillations behave as a matter-like component and can effectively reheat the universe through the decay of the inflaton into relativistic particles. This provides a natural mechanism for the transition from inflationary epoch to the standard hot Big Bang evolution.

Now using  \eq{ephn} the above expression for potential can be rewritten as
\beq\label{vn}
V(N)={H_0}^2{\rm M_{P}}^2\left[3-\frac{2\alpha^2}{(6\alpha \text{N}+\alphaa)^{4/3}}\right] e^{2\alpha\left(1-\frac{1}{\sqrt[3]{6\alpha \text{N}+\alphaa}}\right)}. 
\eeq
So, given an expression for the inflationary equation-of-state one can effectively deduce the associated potential. Consequently, one may argue that Mukhanov parametrization is not indeed a model independent way to study the cosmic inflation. 
Choosing an inflationary  potential directly maps into the underlying high energy physics but with a specific form of the inflationary equation-of-state we might find inflationary observables more easily. So equation-of-state formalism may be seen as bottom-up approach towards cosmic inflation.

%%%%%%%%%%%%%%%%%%%%%%%%%%%%%%%%%%%%%%%%%%%%%%%%%%%%%%%%%%%%%%%%%%%%%%%%%%%%%%%%%%%%%%%%%%%%%%%%%%%%%%%%%%%%%%%%%%%%%%%%%%%%%%%%%%%%%%%%%%%%%%%%%%%%%%%%%%%%%%%%%%%%%%%%%%%%
\section{Lyth Bound}
\begin{figure}%[htb]
	\centerline{\includegraphics[width=12.cm, height=9cm]{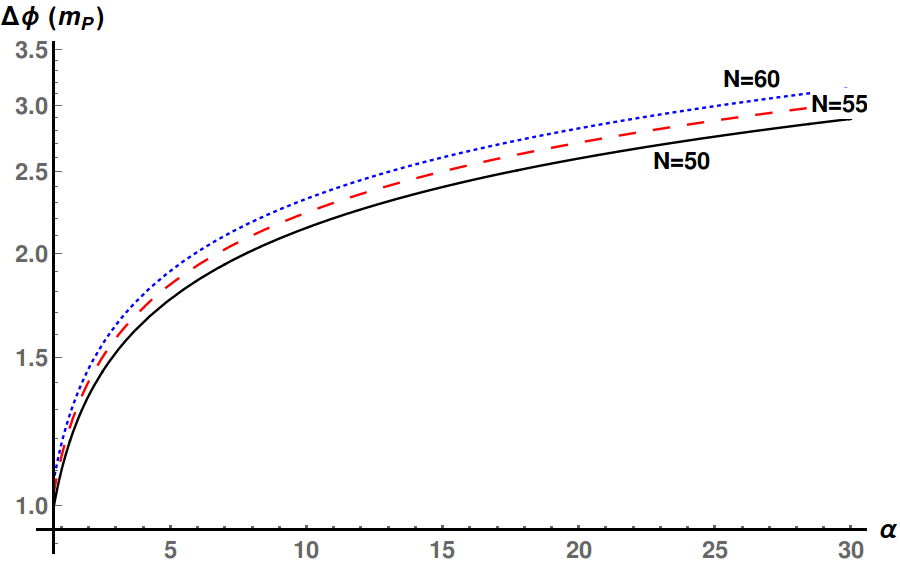}}
	\caption{\label{fig_delta_phi} The variation of  $\Delta \phi$ in the unit of Planck mass, $m_P$,  with the model parameter $\alpha$ for three different values of $N$. }
\end{figure}

The curvature perturbation depends on Hubble Parameter and inflaton, but tensor perturbation depends on Hubble Parameter alone. As a consequence their ratio, tensor-to-scalar ratio, explicitly determines the excursion of inflaton during observable inflation,  known as Lyth bound \cite{lyth1997}, is determined by the following quantity  
\bea
\Delta \phi &\equiv&\phi-\phiend \nonumber\\ &=&\frac{m_P}{\sqrt{8\pi}}\left(\sqrt[3]{6\alpha \text{N}+\alphaa}-\sqrt[3]{\alphaa}\right),
\eea
which is generally expressed in units of Planck mass $m_P$. In \fig{fig_delta_phi} we have  illustrated the variation of  $\Delta \phi$ in the unit of Planck mass for three different values of $\text{N}$. From the figure we see that for this model $\Delta \phi$ is always greater than unity, which implies that quasi-exponential model falls in the wide class of large field model \cite{linde1983}.  
Consequently, quasi-exponential is expected to produce large amount of primordial gravitational waves. But still this production of gravitational waves can be checked and we can bring that down within the present observational limit without any trouble whatsoever.

%%%%%%%%%%%%%%%%%%%%%%%%%%%%%%%%%%%%%%%%%%%%%%%%%%%%%%%%%%%%%%%%%%%%%%%
\section{Inflationary Observables}
Considering $\text{N}$ as new time variable it is possible to express all the inflationary observables in terms of the equation-of-state parameter \cite{mukhanov2013quantum, garcia2014large}, up to the first order in slow-roll parameters which are given by 
\bea
n_{_S}-1 &\simeq& -3\left(1+\omega\right) + \dfrac{d}{d\text{N}}\ln\left(1+\omega\right)\\
\alpha_{_S} &\simeq& 3\dfrac{d}{d\text{N}}\left(1+\omega\right)-\dfrac{d^2}{d\text{N}^2}\ln\left(1+\omega\right)\\
r&\simeq&24\left(1+\omega\right)\\
n_{_T}&\simeq&-3\left(1+\omega\right)\\
\alpha_{_T} &\simeq&3\dfrac{d}{d\text{N}}\left(1+\omega\right).
\eea
The above observable quantities are evaluated  at the time of horizon crossing, i.e. when there are 50-60 e-foldings still left before the end of inflation. In the context of quasi-exponential inflation we find that 
\bea
n_{_S}&\simeq&1-\frac{4\alpha}{(6\alpha \text{N}_{\text{CMB}}+\alphaa)^{4/3}}\left(\alpha+2\sqrt[3]{6\alpha \text{N}_{\text{CMB}}+\alphaa}\right)\\
\alpha_{_S}&\simeq& -\frac{16\alpha^2}{(6\alpha \text{N}_{\text{CMB}}+\alphaa)^{7/3}}\left(2\alpha+3\sqrt[3]{6\alpha \text{N}_{\text{CMB}}+\alphaa}  \right)\\
r&\simeq&\frac{32\alpha^2}{(6\alpha \text{N}_{\text{CMB}}+\alphaa)^{4/3}}\\
n_{_T}&\simeq-&\frac{4\alpha^2}{(6\alpha \text{N}_{\text{CMB}}+\alphaa)^{4/3}}
\eea
where $\text{N}_{\text{CMB}}$ is the number of e-foldings left before the end of inflation. 
To confront with the recent observational data, above quantities are to be evaluated at the time of horizon crossing. So with the specific equation of state parameter one can very easily have the estimate of observable parameters.

%%%%%%%%%%%%%%%%%%%%%%%%%%%%%%%%%%%%%%%%%%%%%%%%%%%%%%%%%%%%%%%%%%%%%%%%%%%%%%%%%%%%%%%%%%%%%%%%%%%%%
\section{Confrontation  with Planck}
The power spectrum of the curvature perturbation up to the first order in slow roll parameters  is given by \cite{liddle1994, kinney2002}
\bea
P_{s}&\simeq&\frac{1}{16\pi^2 \rm M_{P}^4} \bigg[\frac{H(\phi)^2}{H'(\phi)}\bigg]^2_{\phi=\phicmb}\nonumber\\
&=&
\frac{\text{H}_0^2}{16\pi^2\alpha^2\rm M_{P}^2} (6\alpha \text{N}_{\text {CMB}}+\alphaa)^{4/3}
e^{2 \alpha\left(1-\frac{1}{\sqrt[3]{6\alpha \text{N}_{\text {CMB}}+\alphaa}}\right)}
%&=&\frac{2\text{H}_0^2}{\pi^2\rm M_{P}^4} \  \frac{e^{2 \alpha \ -\  {\alpha}^{1/2} \ 2^{-1/4}\ r ^{1/4}  }}{r}
\eea
where $\text{N}_{\text{CMB}}$ is the number of e-foldings still left before the end of inflation when a particular mode crosses the horizon.  
\begin{figure}%[htb]
	\centerline{\includegraphics[width=14.cm, height=9cm]{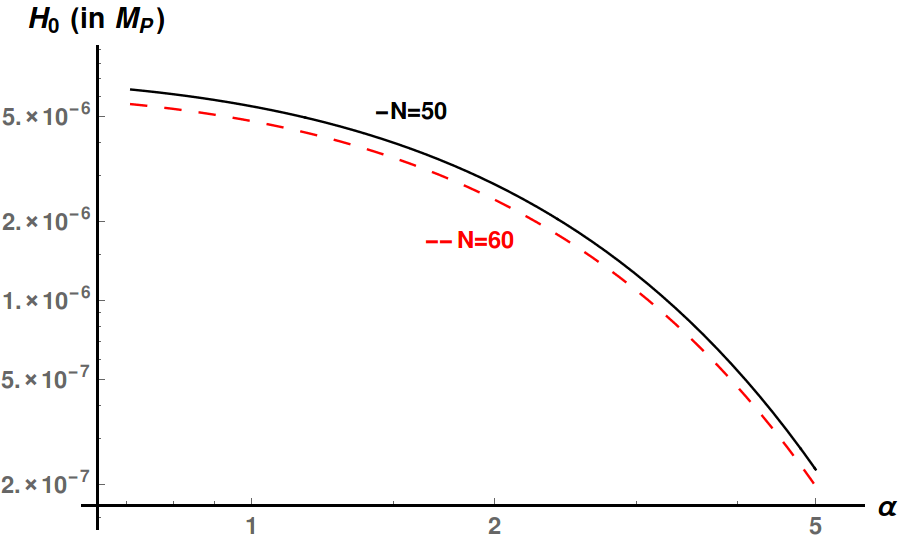}}
	\caption{\label{fig_h0}Variation of $H_0$ in the unit of reduced Planck mass, $M_P$,  with the model parameter $\alpha$ for two different values e-foldings,  $N=50,\ 60$. }
\end{figure}
In \fig{fig_h0} we have shown variation of the inflationary energy scale ($H_0$) with the model parameter for two different values of number of e-foldings, the black solid line represents $N=50$ and red dashed line is for $N=60$. For the above figure we have assumed that the amplitude of inflationary scalar perturbation is $2.12
\times 10^{-9}$ \cite{akrami2020planck, aghanim2020planck}. From the plot we may argue that QEI has the potential to address wide range of inflationary energy scales and hence can easily handle different orders of tensor-to-scalar ratio. Though QEI may not be able to explain tensor-to-scalar ratio $r<0.01$, which  we shall see later.  

\begin{figure}%[htb]
	\centerline{\includegraphics[width=14.cm, height=8cm]{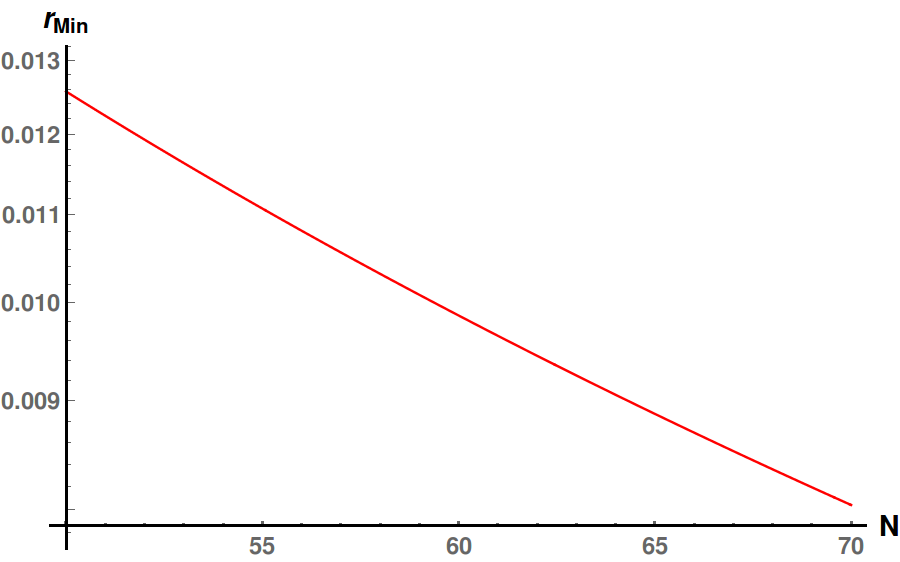}}
	\caption{\label{fig_tensor_min}Variation of lower bound of $r$ with the number of e-folding.}
\end{figure}

The amplitude of the gravitational waves generated during inflation is  expressed by the tensor-to-scalar ratio, which  in QEI is found to be 
\bea
r&\simeq&\frac{32\alpha^2}{(6\alpha \text{N}_{\text{CMB}}+\alphaa)^{4/3}}
\eea
\begin{figure}%[htb]
	\centerline{\includegraphics[width=14.cm, height=8cm]{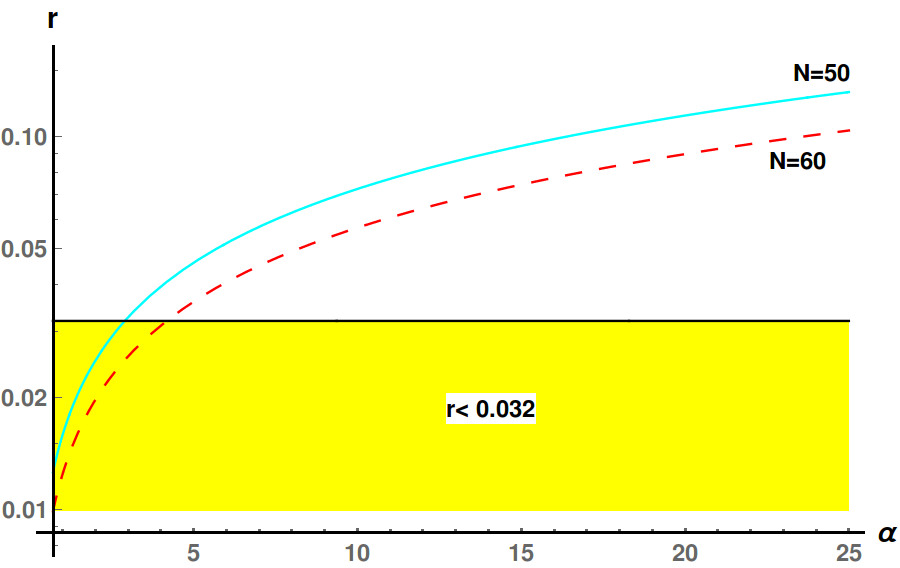}}
	\caption{\label{fig_tensor}Variation of $r$ with the model parameter $\alpha$ for two different values e-foldings,  $N=50,\ 60$. The shaded area corresponds to the  region where $r<0.032$.}
\end{figure} 
In \fig{fig_tensor} we have illustrated  tensor-to-scalar ratio with the model parameter for two different values of number of e-foldings, the solid cyan line represents $N=50$ and red dashed line is for $N=60$. From the recent measurements, the upper-bound on tensor-to-scalar ratio is  $r<0.032$ \cite{tristram2022improved}, which provides the following upper-limit for the model parameter depending on number of e-foldings,  $\alpha\leq 2.90,\ 3.51, \ 4.18$ for 
$\text{N}_{\text{CMB}}=50,\ 55,\ 60$ respectively. So we notice that in  QEI the recent bound on tensor-to-scalar ratio is satisfied for smaller values of the model parameter $\alpha$. Existing observationally viable region has been illustrated via yellow shaded area. From  \fig{fig_tensor} we see that, tensor-to-scalar ratio varies in proportion with the model parameter. This helps us find the lower bound of $r$ as in QEI we have fixed lower bound of $\alpha\geq \frac{1}{\sqrt{2}}$.  In \fig{fig_tensor_min} we have shown variation of the minimum value of tensor-to-scalar ratio with different values of number of e-foldings. From the above figure it is obvious that QEI may not  be able to explain observations with $r< 10^{-2}$. 

\begin{figure}%[htb]
	\centerline{\includegraphics[width=14.cm, height=8cm]{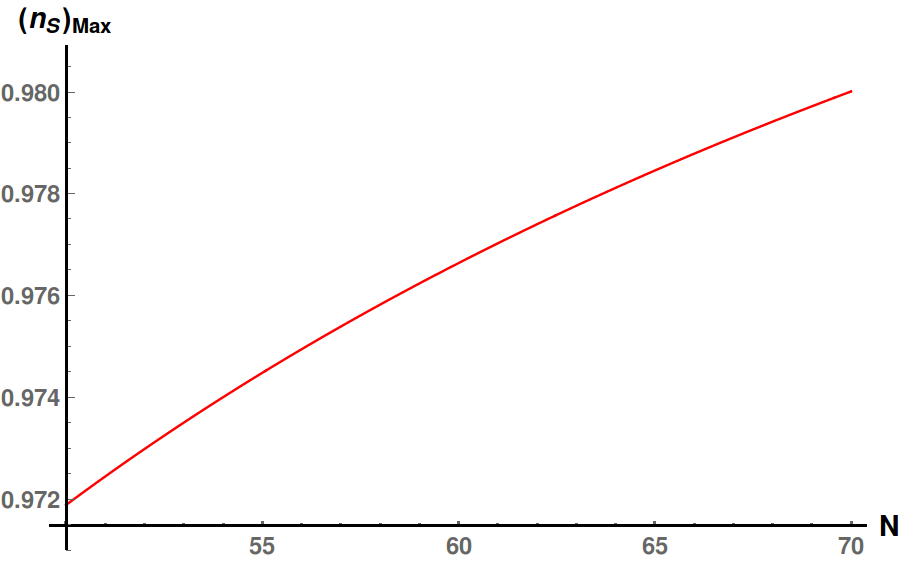}}
	\caption{\label{fig_spectral_max}Variation of upper bound of scalar spectral index with the number of e-folding in quasi-exponential inflation.}
\end{figure}

The scale dependence of the curvature perturbation, is expressed in terms of the spectral index and we have in QEI
\bea
n_{_S}&\simeq&1-\frac{4\alpha}{(6\alpha \text{N}_{\text{CMB}}+\alphaa)^{4/3}}\left(\alpha+2\sqrt[3]{6\alpha \text{N}_{\text{CMB}}+\alphaa}\right). 
\eea
\begin{figure}%[htb]
	\centerline{\includegraphics[width=14.cm, height=8cm]{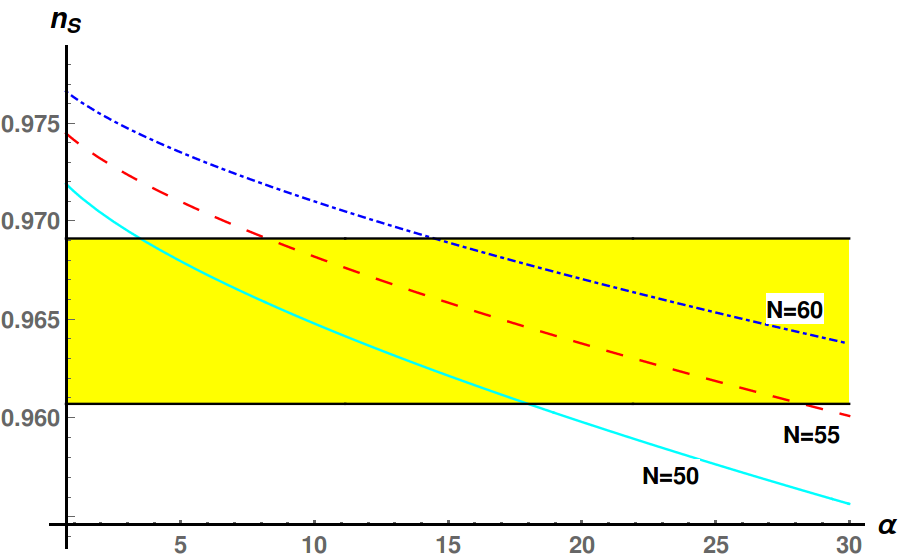}}
	\caption{\label{fig_spectral}Variation of scalar spectral index, $n_{_S}$, with the model parameter $\alpha$ for two different values e-foldings,  $N=50,\ 55, \ 60$. The yellow shaded region corresponds to the 1-$\sigma$ constraint on spectral index from Planck-2018 result \cite{ade2021improved}.}
\end{figure} 
In \fig{fig_spectral} we have plotted the scalar spectral index with the model parameter for three different values of $N$. The yellow shaded region corresponds to the latest 1-$\sigma$ bound  on the spectral index \cite{ade2018constraints,ade2021improved}. From the 1-$\sigma$ constraint  on the scalar spectral index \cite{akrami2020planck, aghanim2020planck}, $0.9607\leq n_{_S}\leq 0.9691$ for Planck TT, TE, EE + lowE + lensing data combination, we find that $3.4781\leq\alpha\leq17.9895, \ 8.21871\leq\alpha\leq 28.2401, \ 14.4859\leq\alpha\leq 40.5177$ for $\text{N}_{\text{CMB}}=50,\ 55, \ 60$ respectively. We notice that constraint on scalar spectral index is satisfied at higher values of the model parameter for QEI. From the plot we also figure out that scalar spectral index varies inversely with the model parameter. With the help of this we provide upper bound of $n_{_S}$ for different values of number of e-foldings. In \fig{fig_spectral_max} we have shown the variation of maximum value achievable for $n_{_S}$ in QEI with the number of e-foldings. We find that QEI can accommodate scalar spectral index up to $0.98$.  

\begin{figure}%[htb]
	\centerline{\includegraphics[width=14.cm, height=10cm]{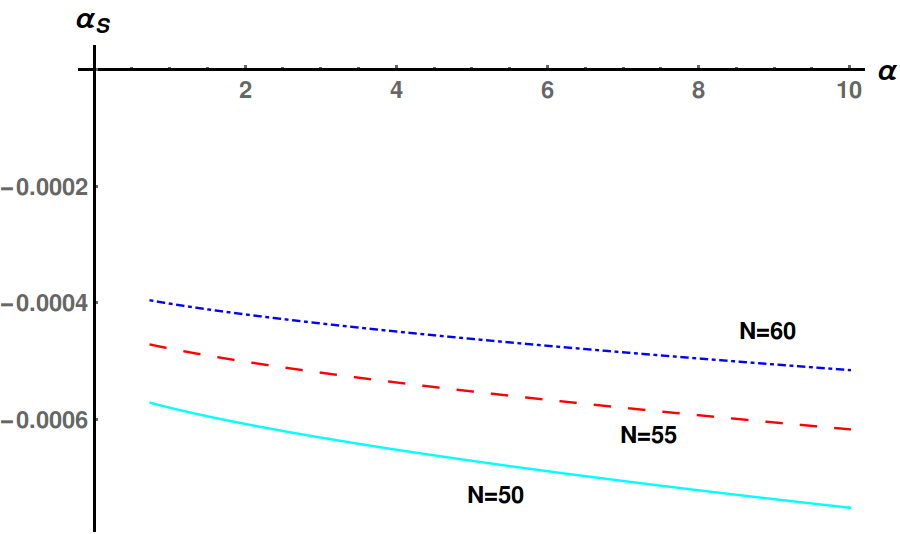}}
	\caption{\label{fig_running}Variation of scalar running, $\alpha_{_S}$, with the model parameter, $\alpha_{_S}$, for three different values of e-foldings,  $N=50,\ 55, \ 60$.}
\end{figure}

The running of spectral index in this case is found to be 
\bea
\alpha_{_S}&\simeq& -\frac{16\alpha^2}{(6\alpha \text{N}_{\text{CMB}}+\alphaa)^{7/3}}\left(2\alpha+3\sqrt[3]{6\alpha \text{N}_{\text{CMB}}+\alphaa}  \right)\label{scalar-running}. 
\eea
Therefore, the scale dependence of the scalar spectral index  is negative definite in QEI, but very small  $\mathcal{O}(10^{-4})$. And hence almost consistent with zero as depicted in \fig{fig_running}.  
%Also from the \eq{scalar-running} it is quite simple to check that for a fixed value of $N_{\text{CMB}}$,  $\alpha_{_S}$ is approximately proportional to  $\alpha^0$, i.e. almost independent of the model parameter which is clearly visible from \fig{fig_running}. While $\alpha_{_S}$  depends on the number of e-foldings as shown in \fig{fig_running}. 

\begin{figure}%[htb]
	\centerline{\includegraphics[width=14.cm, height=10cm]{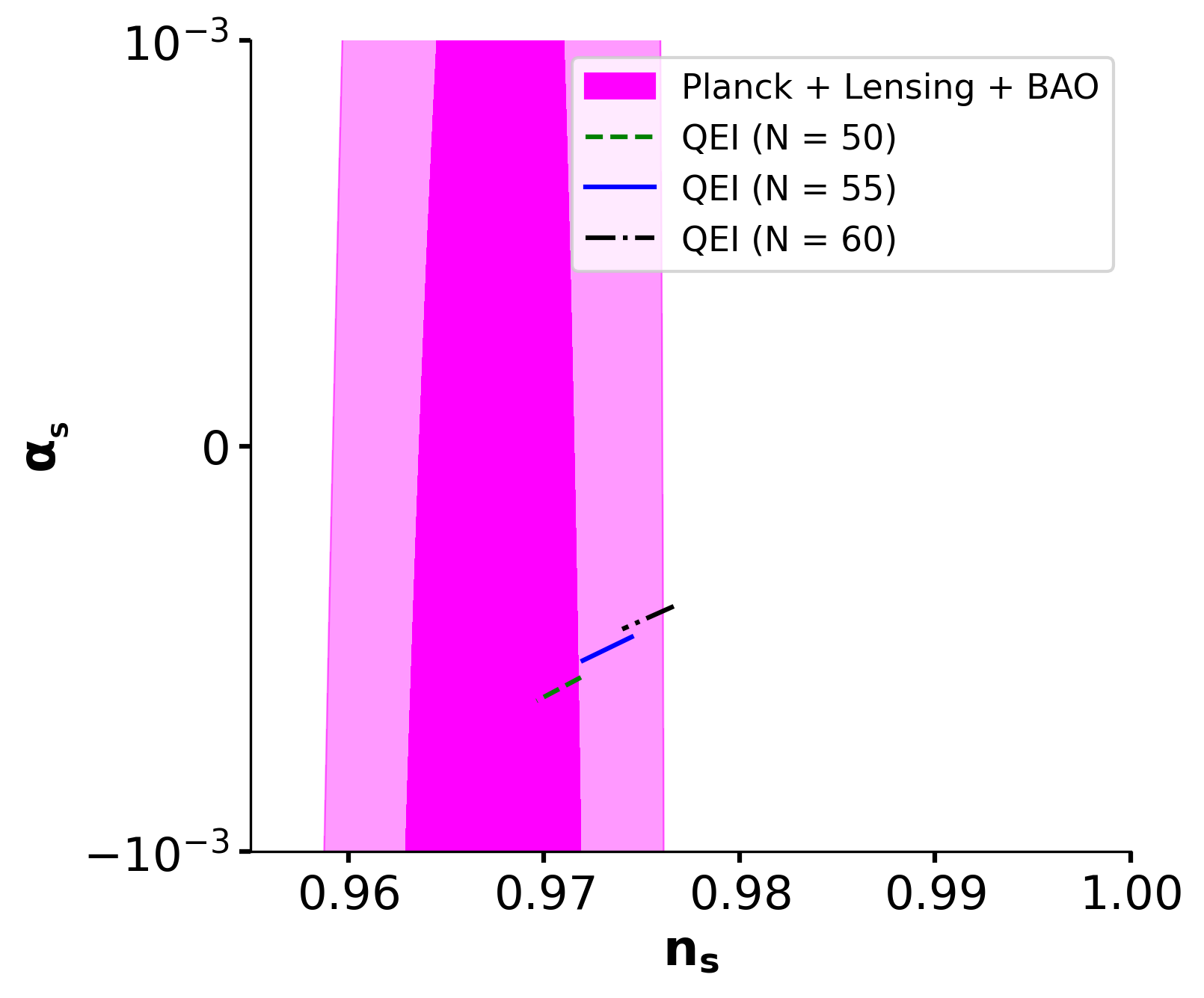}}
	\caption{\label{fig_srunning}Variation of scalar running, $\alpha_{_S}$, with the spectral index, $n_{_S}$, for three different values of e-foldings,  $N=50,\ 55,\ 60$. Marginalized $68\%$ and $95\%$ confidence region in the plane of $\alpha_{_S}-n_{_S}$ from the Planck results \cite{ ade2018constraints, ade2021improved, tristram2024}. The green, blue and black lines represent theoretical predictions from QEI in the  $\alpha_{_S}-n_{_S}$ plane for $N=50, \ 55, \ 60$ respectively, scanned over different values of the model parameter.}
\end{figure}

The Planck-2018 analysis has tighten the constraint on running of the scalar spectral index and 1-$\sigma$ limit is $-0.008\leq \alpha_{_S}\leq0.012$ for Planck TT, TE, EE + lowE + lensing data combination. In \fig{fig_srunning} we have shown variation of $\alpha_{_S}$ with $n_{_S}$ for three values of number of e-foldings. 
It is transparent from the above figure that the running of scalar spectral index as obtained from QEI are within $95\%$ confidence region over a broad parameter range. So, we may conclude that prediction for running of the spectral index from QEI for $\alpha_{_S}$ are in tune with Planck-2018 data.

%$running=−0.0045 ± 0.0067$
%$-0.012<n^{\prime}_{_S}<0.0022$
\begin{figure}%[htb]
	\centerline{\includegraphics[width=14.cm, height=10cm]{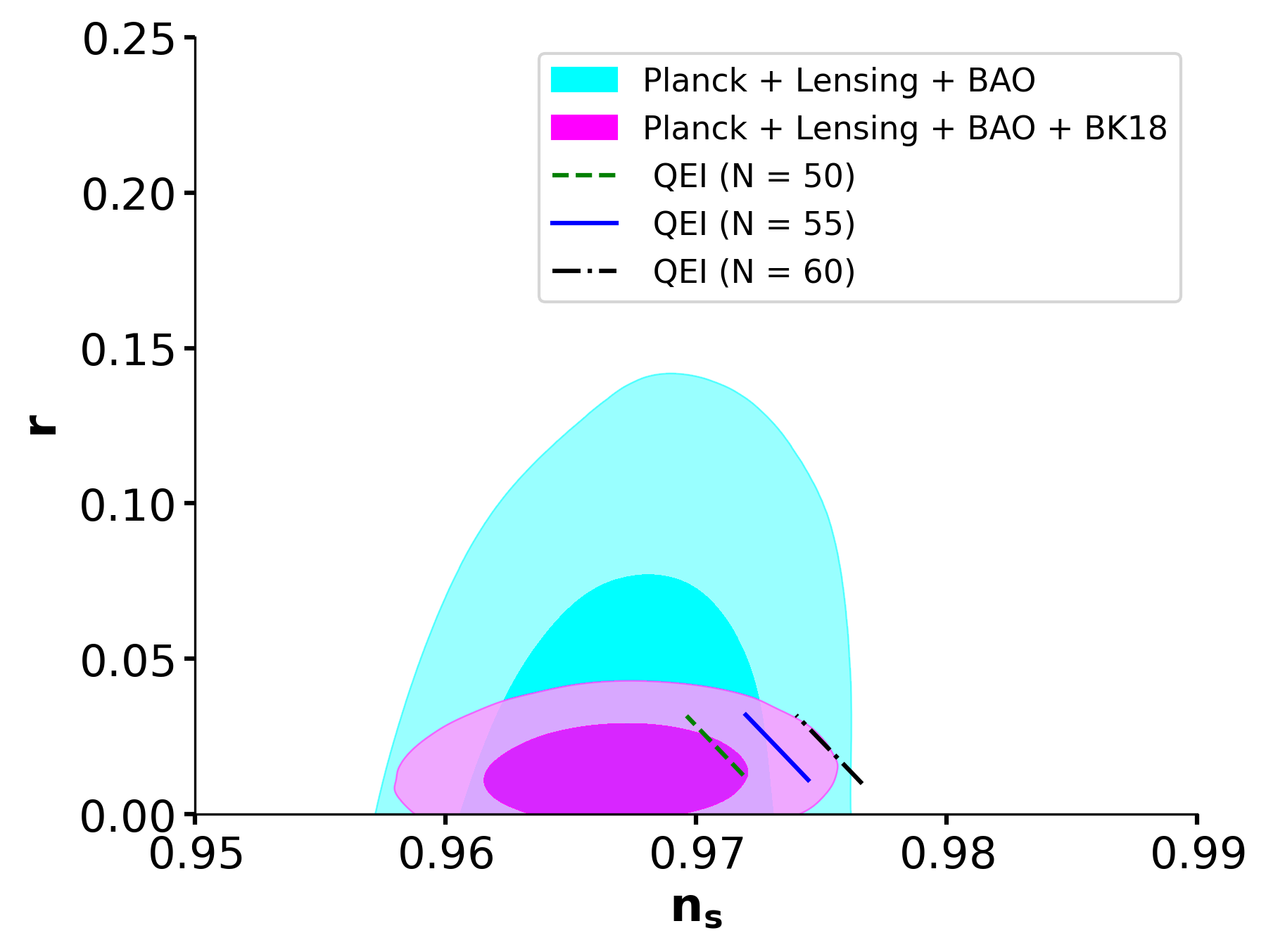}}
	\caption{\label{fig_rns} $68\%$ and $95\%$ confidence contours in the plane of $r-n_{_S}$ from the joint analysis of Planck-2018 and BICEP/Keck 2018 (BK18) data \cite{ade2018constraints,ade2021improved,tristram2022improved, tristram2024}. The dashed-green, solid-blue and dot-dashed black lines represent quasi-exponential model predictions in the  $r-n_{_S}$ plane for $N=50$, $N=55$ and $N=60$ respectively and for different values of the model parameter. }
\end{figure}
In \fig{fig_rns} we have demonstrated our model predictions for  tensor-to-scalar ratio and scalar spectral index for three different values of e-foldings in the $r-n_{_S}$ plane. The shaded region corresponds to   observationally viable $68\%$ and $95\%$ confidence contours from the joint analysis of BICEP/Keck 2018 (BK18) and Planck data \cite{ade2018constraints,ade2021improved,tristram2022improved, tristram2024}. From the figure it is obvious that the prediction from QEI in  $r-n_{_S}$ plane lie well within the current observational bounds. Therefore we may interpret  that QEI does   comply with recent restrictions on  tensor-to-scalar ratio and the scalar spectral index as set by  the joint analysis of Planck and BK18 data. 

\begin{figure}%[htb]
	\centerline{\includegraphics[width=14.cm, height=10cm]{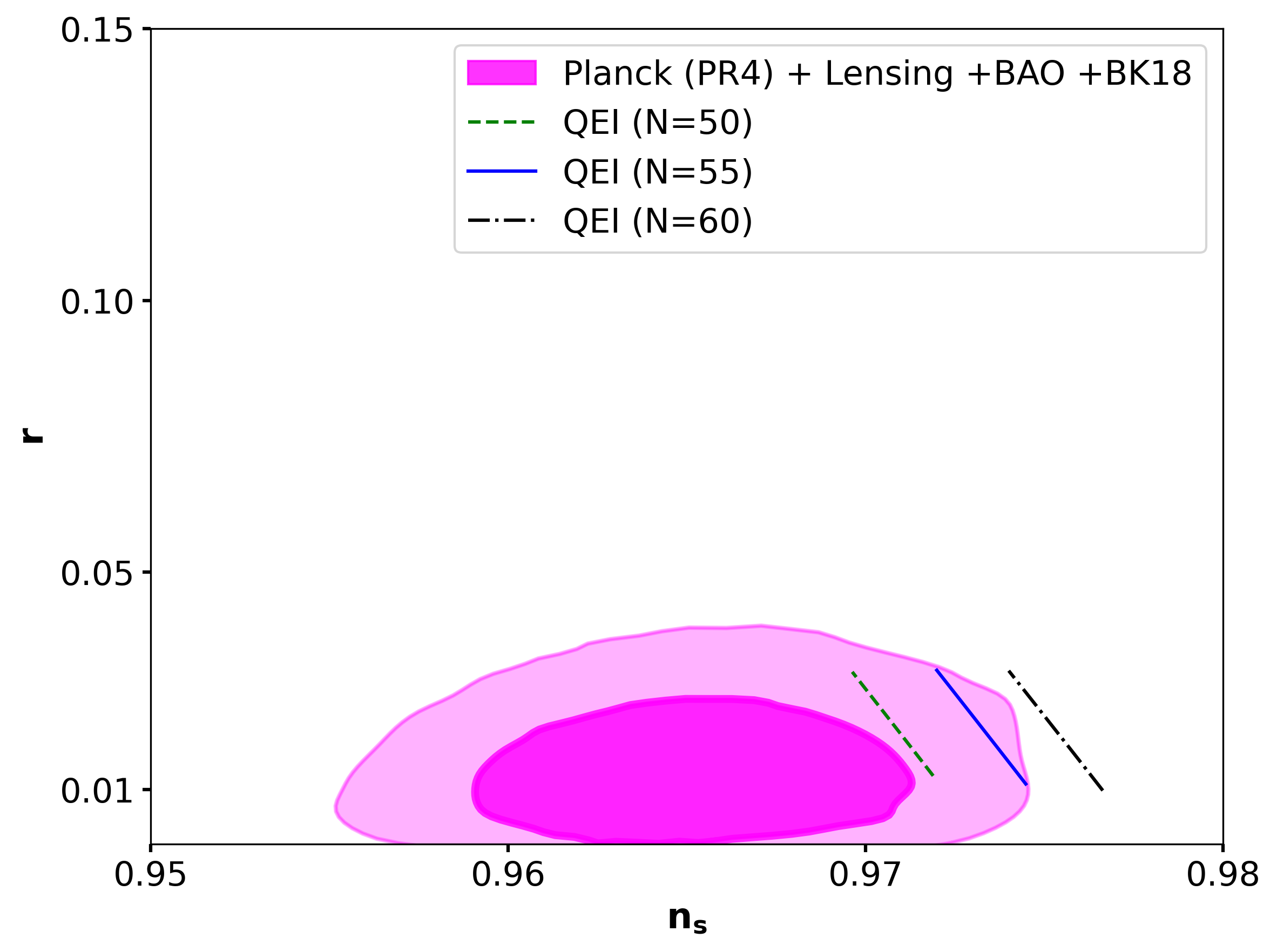}}
	\caption{\label{fig_rns-pr4} Constraints in the $r-n_{_S}$ plane of  from the joint analysis of Planck (PR4) and BICEP/Keck 2018 (BK18) data along with lensing and BAO. This figure has been taken from Ref.\cite{tristram2022improved}. The dashed-green, solid-blue and dot-dashed black lines represent quasi-exponential model predictions in the  $r-n_{_S}$ plane for $N=50$, $N=55$ and $N=60$ respectively and for different values of the model parameter. }
\end{figure}

In \fig{fig_rns-pr4} we have shown constraints in tensor-to-scalar ratio and scalar spectral index in their plane. The constraints are derived by the joint analysis of Planck (PR4) and BK18 data in combination with lensing and BAO. The prediction from quasi-exponential inflation for $r$ and $n_{_S}$ has also been depicted for different values of $N$ by dashed, solid and dot-dashed lines. We see that QEI prediction in the  $r-n_{_S}$ plane lie outside the $68\%$ confidence level but inside $95\%$ contour for $N=50$ and $N=55$.

\section{Confrontation with latest ACT-DR6 Data}
%%%%%%%%%%%%%%%%%%%%%%%%%%%%%%%%%%%%%%%%%%%%%%%%%%%%%%%%%%%%%%%%%%%%%%%%%%%%%%%%%%%%%%%%%%%%%%%%%%%%%%%%%%%%%%%%%%%%%%%%%%%%%%%%%%%%%%%%%%%%%%%%%%%%%%%%%%%%%%%%%%%%%%%%%%%%%%%%%%%%%%%%%%%%%%%%%%%%%%%%%%%%%%%%%%
Very recent data released by Atacama Cosmology Telescope (ACT) \cite{louis2025atacama, calabrese2025atacama} has  shifted the constraint on scalar spectral index towards unity, $n_{_S}=0.9666\pm0.0077$. The combined analysis of Planck joint with BK18 and ACT-DR6  suggested higher value for spectral index, $n_{_S}=0.9709\pm 0.0038$. This value is further increased to $n_{_S}=0.9743\pm 0.0034$ when Planck joint with BK18 and ACT-DR6 are combined with DESI Y1 data \cite{adame2025desi1, adame2025desi}. This is very interesting finding as this indicates that curvature perturbation is more scale invariant than that was found by Planck previously.  In other words, inflationary models which are closer to quasi De-Sitter will now be observationally more preferred. And  quasi-exponential inflation which naturally tends to predict higher value of the scalar spectral index now may be considered to  have the  endurance and vision to come out as the winner. 

\begin{figure}%[htb]
	\centerline{\includegraphics[width=14.cm, height=9cm]{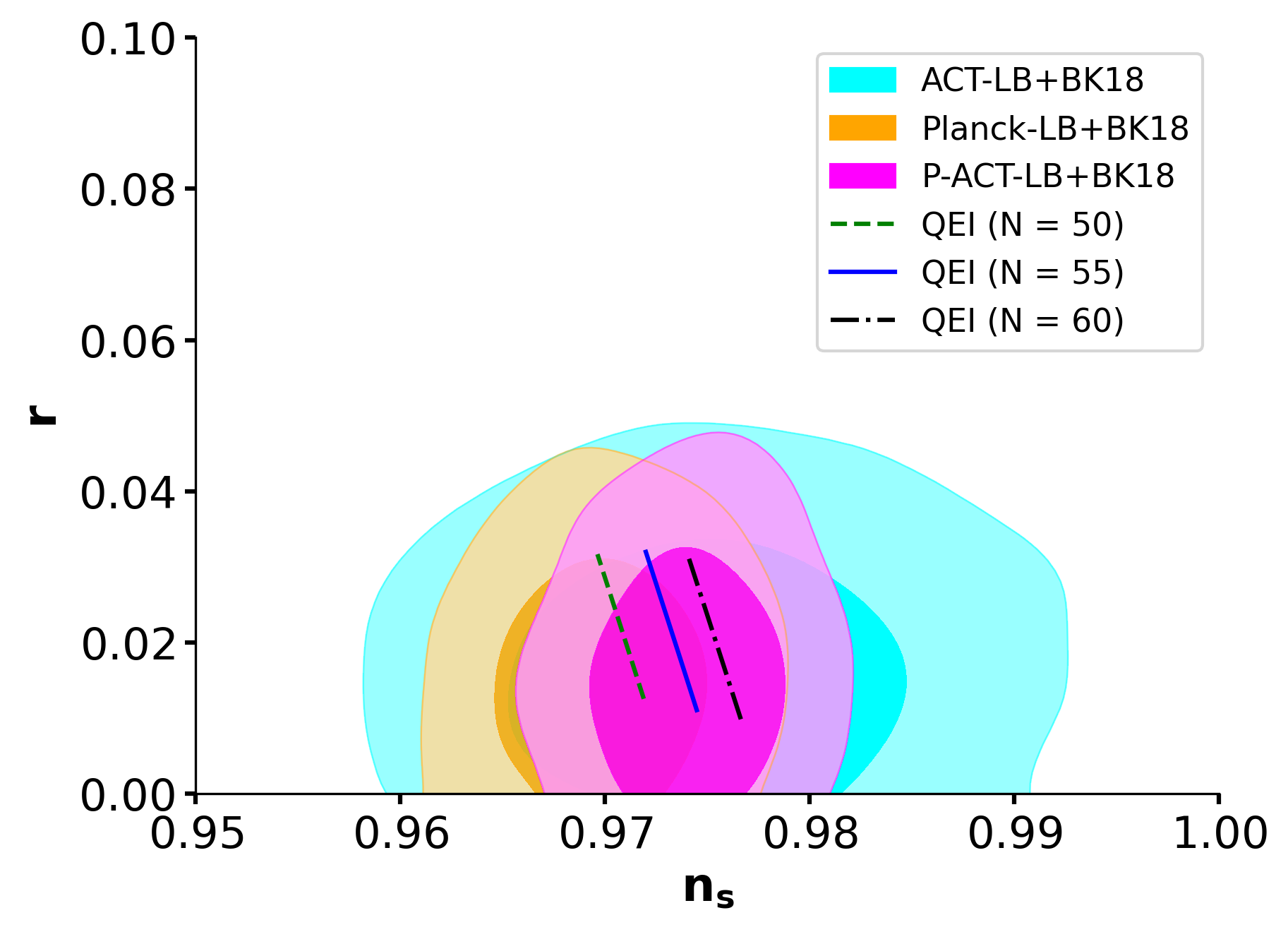}}
	\caption{\label{fig_rnsactdesi}Marginalized joint confidence contours of  $68\%$ and $95\%$ in the space of tensor-to-scalar ratio, $r$, and scalar spectral index, $n_{_S}$. The constraint on $r$ derived from BICEP2/Keck or BK18 data \cite{ade2018constraints}, while constraint on  $n_{_S}$ is driven by the  combination of Planck, ACT-DR6 and DESI Y1 data. The dashed green, solid blue and dot-dashed black  lines are the $n_{_S}-r$	relations as obtained	from  QEI for $N=50, \ 55, 60$ respectively.}
\end{figure}

The quasi-exponential model of inflation renders excellent fit to the outcome of combined analysis of Planck joint with BK18, ACT-DR6  and DESI Y1 data, as depicted in \fig{fig_rnsactdesi}. The marginalized joint confidence contours of  $68\%$ and $95\%$  in $r-n_{_S}$ plane is from the joint analysis of Planck, ACT-DR6 and DESI Y1 data along with the present constraint on primordial gravitational waves, $r<0.032$, from BK18. In this case we find that the upper limit of $\alpha$ is  $\alpha\leq 3.51$ for $N=55$.  The lower limit, independent of number of e-foldings, has been fixed earlier to $\alpha \geq\frac{1}{\sqrt{2}}$. 
From the figure we can conclude that the inflationary predictions for tensor-to-scalar ratio and spectral index from quasi-exponential inflation are in excellent agreement with the combined analysis of ACT-DR6, Planck joint with BK18 and  DESI Y1 data. 

\begin{figure}%[htb]
	\centerline{\includegraphics[width=14.cm, height=9cm]{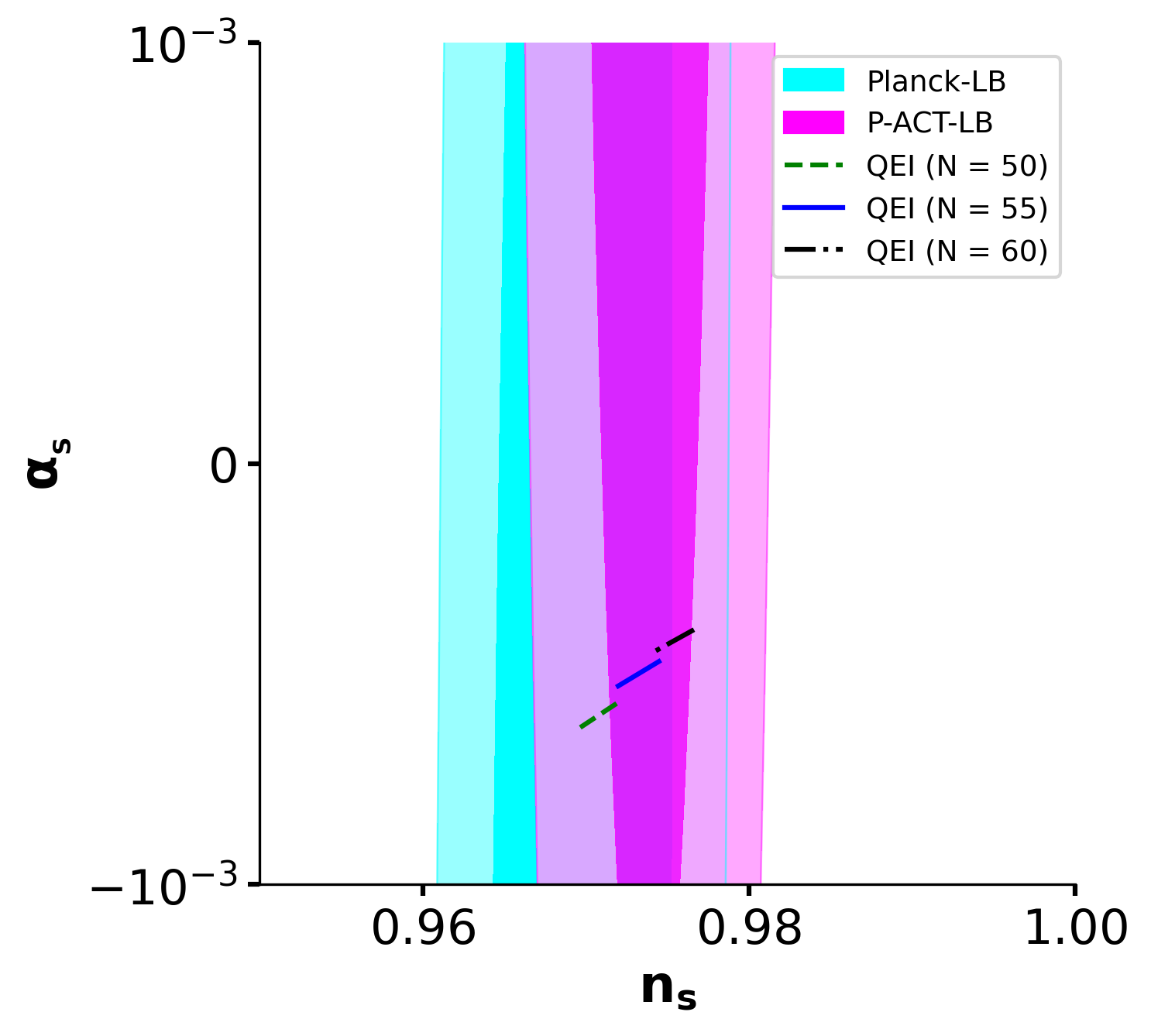}}
	\caption{\label{fig_srunning_pactlb}Marginalized  $68\%$ and $95\%$ confidence level constraints in running of the spectral index, $\alpha_{_S}$ and  scalar spectral index, $n_{_S}$, for the  combination of Planck,  ACT-DR6 and DESI Y1 data. The dashed green, solid blue and dot-dashed black  lines are the $n_{_S}-\alpha_{_S}$	relations as obtained	from  QEI for $N=50, \ 55, 60$ respectively.}
\end{figure}
In \fig{fig_srunning_pactlb} we have shown the prediction of running of the scalar spectral index with $n_{_S}$ itself from QEI in plane of $n_{_S}-\alpha_{_S}$. We find that QEI is in tune with the latest constraints from the combined analysis of ACT-DR6, Planck joint with BK18 and DESI Y1 data. 

%%%%%%%%%%%%%%%%%%%%%%%%%%%%%%%%%%%%%%%%%%%%%%%%%%%%%%%%%%%%%%%%%%%%%%
%%%%%%%%%%%%%%%%%%%%%%%%%%%%%%%%%%%%%%%%%%%%%%%%%%%%%%%%%%%%%%%%%%%%%%
\section{Quasi-Exponential Inflation and Future CMB Missions}
Before moving to the forecast analysis with future CMB missions, we outline the viable region in the $r$--$n_{_S}$ space where QEI holds.
\begin{figure}
	\centerline{\includegraphics[width=14.cm, height=8cm]{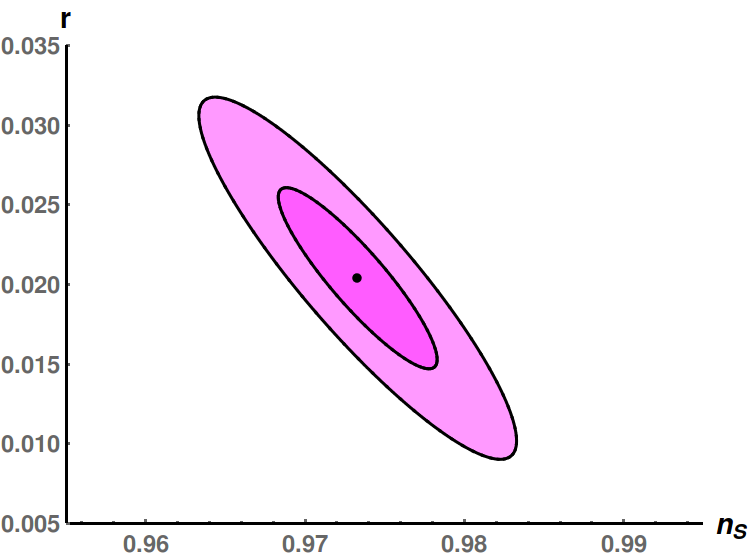}}
	\caption{\label{fig_rns-model} The 1-$\sigma$ and 2-$\sigma$  confidence ellipses in the $r$--$n_{_S}$ plane derived from QEI model. The ellipses are computed from the covariance of $n_{_S}$ and $r$ evaluated over the allowed ranges of the  model parameter $\alpha$ and varying $N$ between $50$ and $60$.  The black point denotes the mean prediction of the model. These contours represent the intrinsic spread of model predictions and indicate the viable region in the $r$--$n_{_S}$ parameter space. }
\end{figure}
In order to visualize the range of predictions from QEI model, we construct confidence contours in the $r$--$n_{_S}$ plane based on the statistical distribution of $n_{_S}$ and $r$ values from their expressions in QEI model. The resulting $1-\sigma$ and $2-\sigma$ ellipses, shown in \fig{fig_rns-model}, delineate the viable region of the model. The black point corresponds to the mean prediction, while the shaded regions represent the theoretical spread arising from variations in $\alpha$ and $N$. The contour is obtained by varying $N$ between 50 and 60, with model parameter restricted to $\alpha \geq 1/\sqrt{2}$ and the upper limit constrained by the observational bound $r < 0.032$. The plot clearly indicates that the QEI model fails to account for tensor-to-scalar ratios below $r = 0.01$, while remaining consistent with the wide range of scalar spectral index.

LiteBIRD, the Lite (Light) satellite for the study of B-mode polarization and Inflation from cosmic background Radiation Detection, is a space mission for primordial cosmology and fundamental physics. LiteBIRD is promising to detect  
tensor-to-scalar ratio with uncertainty of $\sigma( r)\sim0.001$ \cite{litebird2023probing, ghigna2024litebird}. Detection of gravitational waves at 3-$\sigma$ level by LiteBIRD would  imply $r>0.003$. And non-detection of CMB B-mode polarization, LiteBIRD will set an upper bound on the amplitude of primordial gravitational waves $r<0.002$. 

\begin{figure}
	\centering
	\begin{subfigure}{.5\textwidth}
		\centering
		\includegraphics[width=.96\linewidth,height=6.5cm]{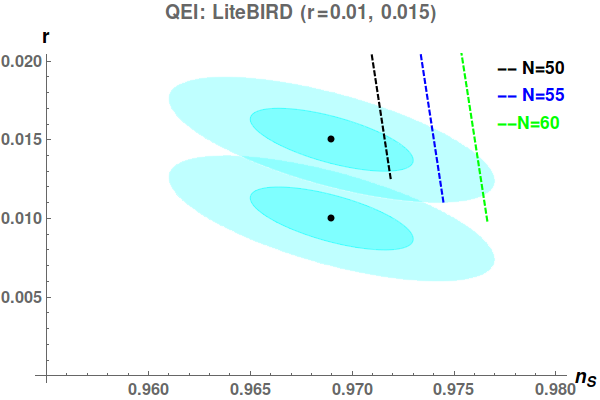}
		\caption{$n_{_S}=0.9690$}
		%\label{fig:sub1}
	\end{subfigure}%
	\begin{subfigure}{.5\textwidth}
		\centering
		\includegraphics[width=.96\linewidth,height=6.5cm]{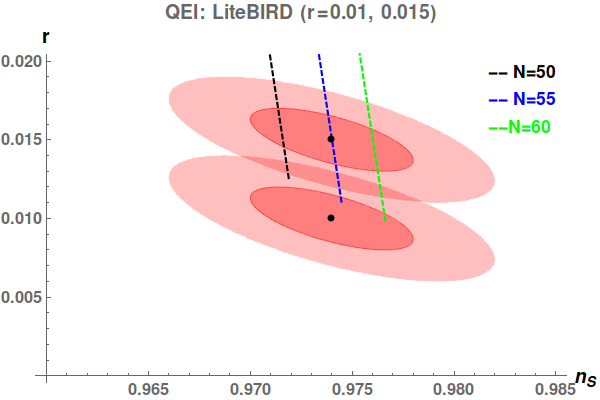}
		\caption{$n_{_S}=0.9743$}
		%\label{fig:sub2}
	\end{subfigure}
	\caption{Forecasts for joint constraints on the tensor-to-scalar ratio and the scalar spectral index for a fiducial model with $r=0.01, \ 0.015$ and $n_{_S}=0.9690\ ({\rm Left Panel}),\ n_{_S}= 0.9743\ ({\rm Right Panel})$. The uncertainties are derived assuming the proposed sensitivity of futuristic space mission LiteBIRD. Dashed black, blue and green lines represent prediction from quasi-exponential model for $N=50$ and $N=60$ respectively. The black dots at centre represent fiducial model.	}
	\label{fig_rnslitebird}
\end{figure}

\begin{figure}%[htb]
	\centerline{\includegraphics[width=12.cm, height=9cm]{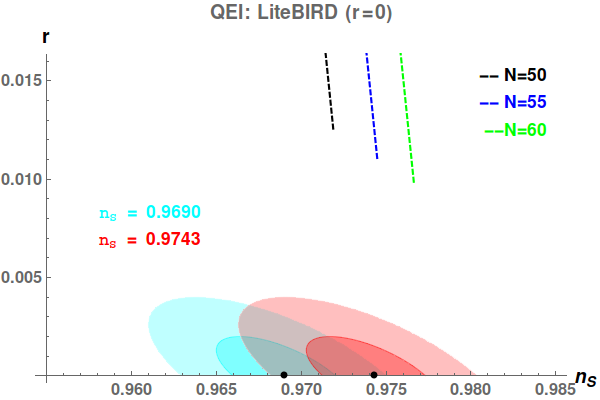}}
	\caption{\label{fig_rnslitebirdn}LiteBIRD constraints in $r-n_{_S}$ plane for a fiducial model with $r=0$ and spectral index $n_{_S}=0.9690,\ 0.9743$. The constraints on $r$ and $n_{_S}$ are driven by the expected sensitivity from the LiteBIRD experiment. Black, blue and green dashed lines represent expected $r-n_{_S}$ from QEI for  $N=50,\ 55, \ 60$ respectively. }
\end{figure}

In \fig{fig_rnslitebird} we have shown forecasted $68\%$ and $95\%$ confidence contours in the $r-n_{_S}$ plane, for a fiducial model with $r=0.01, \ 0.015$ and $n_{_S}=0.9690$ (Left Panel) and  $n_{_S}=0.9743$ (Right Panel). The shaded  region corresponds to the statistical uncertainties, assuming Gaussian likelihoods centred at the fiducial model, as expected from LiteBIRD.  Black, blue and green dashed  lines represent prediction from quasi-exponential model for three different number of e-foldings and model parameter $\alpha$ over an interval. The fiducial model has been marked with a black dot. From the plot we may conclude that  prediction for scalar spectral index and tensor-to-scalar ratio from QEI are in excellent agreement with futuristic LiteBIRD mission for $r=0.01$ and $0.015$.  Not only that QEI will be able to provide wide range of parameter values within which model prediction would be in  tune with the above futuristic CMB mission.
However, if LiteBIRD fails to detect large scale CMB B-mode polarization, then the upper-limit of tensor-to-scalar ratio will be pushed back to $r<0.002$ and in that case quasi-exponential inflation will be ruled out as depicted in \fig{fig_rnslitebirdn}. 

%%%%%%%%%%%%%%%%%%%%%%%%%%%%%%%%%%%%% CMB S4 %%%%%%%%%%%%%%%%%%%%%%%%%%%%%%%%%%%%
The forthcoming ground based CMB-S4 \cite{abazajian2016cmb, abazajian2022cmbS4} mission  is anticipated to detect inflationary gravitational waves provided $r>0.003$ or will set an upper-bound $r<0.001$ \cite{abazajian2022cmbS4,belkner2024cmbs4} in the absence of detection. 
\begin{figure}
	\centering
	\begin{subfigure}{.5\textwidth}
		\centering
		\includegraphics[width=.96\linewidth,height=6.5cm]{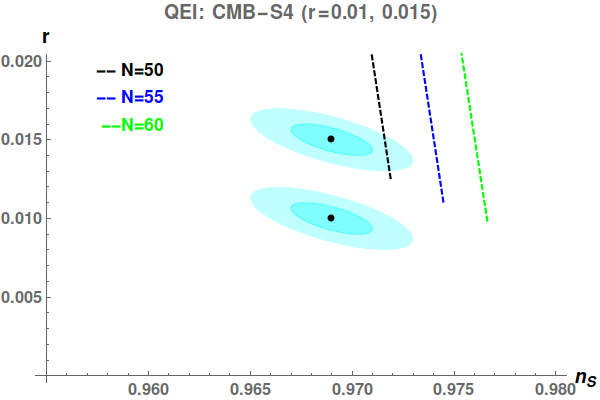}
		\caption{$n_{_S}=0.9690$}
		%\label{fig:sub1}
	\end{subfigure}%
	\begin{subfigure}{.5\textwidth}
		\centering
		\includegraphics[width=.96\linewidth,height=6.5cm]{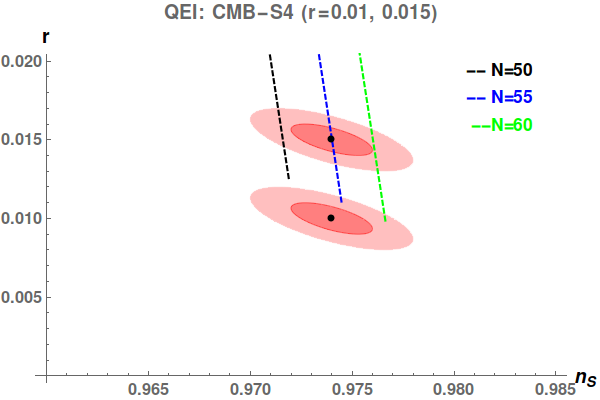}
		\caption{$n_{_S}=0.9743$}
		%\label{fig:sub2}
	\end{subfigure}
	\caption{Marginalized $68\%$ and $95\%$ contours in the $r-n_{_S}$ plane for a fiducial model with $r=0.01, \ 0.015$ and $n_{_S}=0.9690\ ({\rm Left Panel}),\ n_{_S}= 0.9743\ ({\rm Right Panel})$. The shaded  regions corresponds to the statistical uncertainties, assuming Gaussian likelihoods as proposed by CMB-S4. Dashed black, blue and green lines represent prediction from quasi-exponential model for $N=50,\ 55, \ 60$ respectively. The black dots at centre represents fiducial model.	}
	\label{fig_rnss4}
\end{figure}
\begin{figure}%[htb]
	\centerline{\includegraphics[width=12.cm, height=9cm]{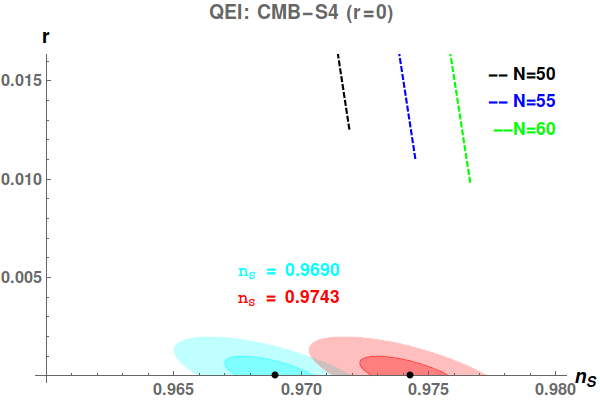}}
	\caption{\label{fig_rnss4n} Forecast of CMB-S4 constraints in the $n_{_S}-r$ plane for a fiducial model with $r = 0$ and  $n_{_S}=0.9690,\ 0.9743$. The constraints are derived from the expected CMB-S4 sensitivity. Mean values of $n_{_S}$ is taken from latest Planck result\cite{tristram2024} and the combined analysis of  ACT-DR6, Planck joint with BK18 and DESI-Y1 data. Dashed black, blue and green lines represent prediction from quasi-exponential model for $N=50, \ 55, \ 60$.}
\end{figure}
In \fig{fig_rnss4} we have illustrated variation of $r$ with $n_{_S}$ for QEI in the $r-n_{_S}$ plane. The shaded area is the forecasted   $68\%$ and $95\%$ confidence region for a fiducial model with $r=0.01,\ 0.015$. Statistical uncertainties are calculated assuming Gaussian likelihoods centred at the fiducial model, expected from futuristic ground based  CMB-S4 mission. We see that QEI prediction for tensor-to-scalar ratio and scalar spectral index are in accord with the futuristic CMB-S4 mission.  A positive detection of tensor-to-scalar ratio by CMB-S4 with r<0.01 can also rule out QEI as shown in \fig{fig_rnss4n}. 

\begin{figure}%[htb]
	\centerline{\includegraphics[width=12.cm, height=9cm]{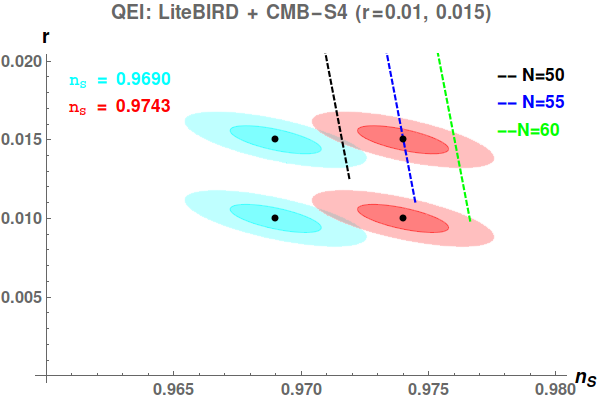}}
	\caption{\label{fig_LBS4-PACT}Marginalized $68\%$ and $95\%$ confidence contours in the $r-n_{_S}$ plane, for a fiducial model with $r=0.01, \ 0.015$ and $n_{_S}=0.9690, \ 0.9743$, assuming joint CMB-S4 + LiteBIRD like experimental setup. Dashed black, blue and green lines represent prediction from the quasi-exponential model for $N=50, \ 55, \ 60$ and $N=60$ respectively. The black dots at centre represent fiducial model.}
\end{figure}

%%%%%%%%%%%%%%%%%%%%%%%%%%%%%%%%%%%%%%%%%%%% LiteBIRD + CMB-S4 %%%%%%%%%%%%%%%%%%%%
In \fig{fig_LBS4-PACT} we have  forecasted marginalized $68\%$ and $95\%$ confidence region in the $r-n_{_S}$ plane. The constraints  are derived assuming the expected sensitivity of the joint analysis of CMB-S4 and LiteBIRD for a fiducial model with $r=0.01, \ 0.015$ and $n_{_S}=0.9690,\ 0.9743$. In this case we found that prediction from QEI is consistent with $n_{_S}=0.9743$, however in order to comply with  $n_{_S}=0.9690$, we may need smaller value for the number e-foldings. 

\begin{figure}%[htb]
	\centerline{\includegraphics[width=12.cm, height=9cm]{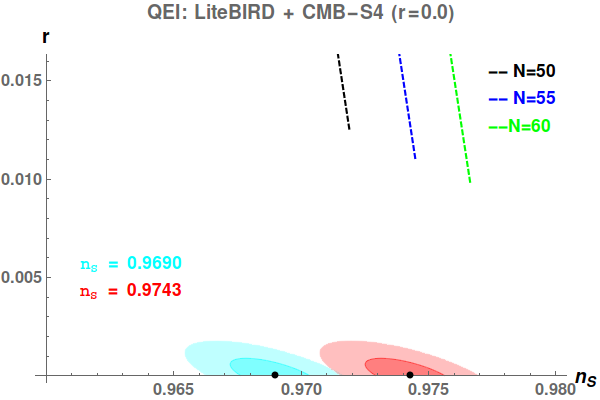}}
	\caption{\label{fig_LBS4-PACT0}Marginalized $68\%$ and $95\%$ confidence contours in the $r-n_{_S}$ plane, for a fiducial model with $r=0.0$ and $n_{_S}=0.9690, \ 0.9743$, assuming CMB-S4 like experimental setup in combination with LiteBIRD. Dashed black, blue and green lines represent the relation between $r-n_{_S}$ from quasi-exponential model for $N=50$ and $N=60$ respectively. The black dots at centre denote fiducial model. }
\end{figure}
In \fig{fig_LBS4-PACT0} we present forecasted marginalized $68\%$ and $95\%$ contours in the $r-n_{_S}$ plane from a joint analysis of CMB-S4 like experimental setup combined with LiteBIRD for a fiducial model with $r=0.0$ and $n_{_S}=0.9690,\ 0.9743$. In this case we found that prediction from QEI are not in good harmony with the forecasted constraints, which may lead to the rejection of QEI model.

Forthcoming CMB mission Simons Observatory (SO) \cite{SimonsObservatory:2018}, a next-generation ground based CMB mission, is aiming to detect primordial gravity waves provided $r\geq0.01$ with sensitivity of $\sigma (r)=0.003$. Further, Simons Observatory will halve the uncertainty in scale  dependence of  scalar curvature perturbation, $\sigma(n_{_S})=0.002$, when used in combination with Planck result.     
\begin{figure}%[htb]
	\centerline{\includegraphics[width=12.cm, height=8.5cm]{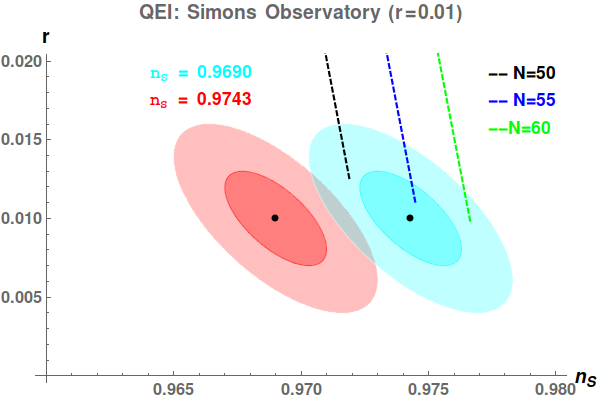}}
	\caption{\label{fig_SO-01}Marginalized $68\%$ and $95\%$ confidence contours in the $r-n_{_S}$ plane, for a fiducial model with $r=0.01$ and $n_{_S}=0.9690, \ 0.9743$, assuming Simons Observatory like experimental setup. Dashed black, blue and green lines represent the relation between $r-n_{_S}$ from quasi-exponential model for $N=50$, $N=55$ and $N=60$ respectively. The black dots at centre denote fiducial model. }
\end{figure}
\begin{figure}%[htb]
\centerline{\includegraphics[width=12.cm, height=9cm]{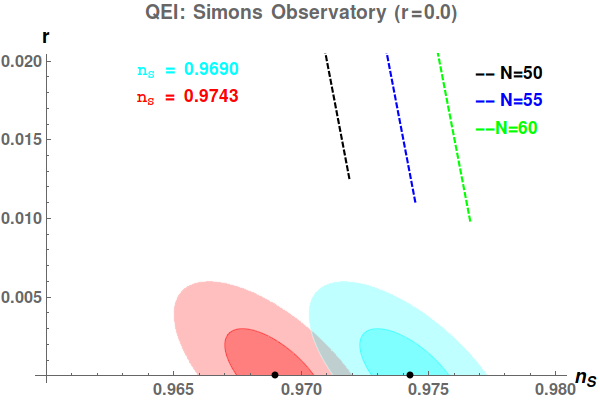}}
\caption{\label{fig_SO-00}Marginalized $68\%$ and $95\%$ confidence contours in the $r-n_{_S}$ plane, for a fiducial model with $r=0.0$ and $n_{_S}=0.9690, \ 0.9743$, assuming Simons Observatory like experimental setup. Dashed black, blue and green lines represent the relation between $r-n_{_S}$ from quasi-exponential model for $N=50$, $N=55$ and $N=60$ respectively. The black dots at centre denote fiducial model. }
\end{figure}
In \fig{fig_SO-01} we present forecasted marginalized $68\%$ and $95\%$ contours in the $r-n_{_S}$ plane for SO like experimental setup for a fiducial model with $r=0.01$ and $n_{_S}=0.9690,\ 0.9743$. In this case we found that prediction from QEI are in  good agreement for $n_{_S}=0.9743$ but not in tune for $n_{_S}=0.9690$. 
In \fig{fig_SO-00} we have forecasted the same but now for $r=0.0$. In this case we find that non-detection of primordial gravity waves by SO will direct the QEI model towards its rejection.

%%%%%%%%%%%%%%%%%%%%%%%%%%%%%%%%%%%%%%%%%%%%%%%%%%%%%%%%%%%%%%%%%%%%%%%%%%%%%%%%%%%%%%%%%%%%%%%%%%%%%%%%%%%%%%%%%%%%%%%%%%%%%%%%%%%%%%%%%%%%%%%%%%%%%%%%%%%%%%%%%%%%%%%%%%%%%%
\section{CONCLUSION}
In this article we have confronted quasi-exponential inflation with the latest ACT-DR6 along with Planck joint with BK18 and futuristic CMB missions in the likes of LiteBIRD, Simons Observatory and CMB-S4 employing Hamilton-Jacobi formulation following Mukhanov parametrization of inflationary equation-of-state. The model having a single free parameter does have a strong agreement with recent observations. The model parameter is constrained by estimating and confronting major inflationary observables   with observational data of late. 

We find that quasi-exponential model can trace wide range of tensor-to-scalar ratio up to $\mathcal{O}({10^{-2}})$. Not only that the scalar spectral index and its running  are also in tune  with Planck data and with combined analysis of ACT-DR6, DESI-Y1 and Planck joint with BK18 data. 

In addition to that, quasi-exponential inflation provides an excellent fit to the latest ACT-DR6 as this data favours higher value for scalar spectral index than Planck prediction for the same. When we consider the combination of ACT-DR6, Planck joint with BK18  along with DESI-Y1 data, we find that the model under consideration does have a observationally viable region for a wide rang of values of the model parameter. This also helps us  put stringent constrain on the  model parameter $\frac{1}{\sqrt{2}}\leq\alpha\leq 3.51$ for $N=55$.  

We have further confronted quasi-exponential model with the upcoming CMB missions, LiteBIRD, SO and CMB-S4. The most fascinating  aspect of our analysis is that quasi-exponential model of inflation will be in tune with the outcomes of LiteBIRD, SO and CMB-S4 provided primordial gravitational waves has been detected by them with $r\geq0.01$. However, if LiteBIRD, SO or CMB-S4 fails to detect any primordial tensor perturbation or tensor-to-scalar ratio is below $\mathcal{O}({10^{-2}})$ then quasi-exponential model will be rejected. 

We have also tested quasi-exponential model with the forecasted  marginalized $68\%$ and $95\%$ confidence contours in the $r-n_{_S}$ plane, for a fiducial model with $r=0.01, \ 0.015$ and $n_{_S}=0.9743, \ 0.9690$. The constraints are derived from  the expected sensitivity of  CMB-S4 in combination with LiteBIRD. Here also we find QEI is in excellent agreement for $n_{_S}=0.9743$, but falls outside the $95\%$ confidence  level for $n_{_S}=0.9690$ and $r=0.01$. 

Finally, we have confronted quasi-exponential inflation with the upcoming ground based next-generation CMB mission Simons Observatory. In this case too we found an excellent fit provided $r\geq0.01$ for $n_{_S}=0.9743$. However, non-detection of primordial gravitational waves by Simons Observatory will summarily reject  quasi-exponential inflation.

In a nutshell inflationary predictions from quasi-exponential inflation are in tune with the latest observations and also with the forecast from futuristic CMB missions in the likes of CMB-S4,  LiteBIRD and Simons Observatory. We also observe that the QEI model fits the data better for higher values of the scalar spectral index, which indicates a preference toward scale invariant scalar curvature perturbations. We have already seen from  \fig{fig_rns-pr4} that prediction from QEI in $r-n_{_S}$ plane lies outside the $1-\sigma$ region due to  its relatively high value of the scalar spectral index. However, as depicted in \fig{fig_rns-model}, the QEI model may also accommodate the observations over a wide range of scalar spectral indices when the model parameter is varied within its allowed range, along with the number of e-folds in the interval $50 \leq N \leq 60$. 
Last but not the least, QEI prediction for tensor-to-scalar ratio is $r\geq 0.01$ which is well within the reach of each of  Simons Observatory, LiteBIRD and CMB-S4. Consequently, sooner or later quasi-exponential inflation will go through the real observational trial.

%%%%%%%%%%%%%%%%%%%%%%%%%%%%%%%%%%%%%%%%%%%%%%%%%%%%%%%%%%%%%%%%%%%%%%%%%%%%
\section*{Acknowledgment}
I would like to sincerely thank anonymous reviewers for their critical and constructive suggestions on the first version of this work.  Which help me a lot to improve the quality of this paper. 
%%%%%%%%%%%%%%%%%%%%%%%%%%%%%%%%%%%%%%%%%%%%%%%%%%%%%%%%%%%%%%%%%%%%%%%%%%%%%%%%%%%%%%%%%%%%%%%%%%%%%%%%%%%%%%% REFERENCES%%%%%%%%%%%%%%%%%%%%%%%%%%%%%%%%%
%\bibliographystyle{unsrt}
%\bibliography{references}

\begingroup
\bibliographystyle{unsrt}
\bibliography{references}
\endgroup

\end{document}